\documentstyle[prd,aps,epsfig]{revtex}
\begin{document}
%
%
\def\ov{\over}
\def\l{\left}
\def\r{\right}
\def\be{\begin{equation}}
\def\ee{\end{equation}}
\draft
\title{A relativistic formalism to compute quasi-equilibrium configurations \\
of non-synchronized neutron star binaries}
\author{Silvano Bonazzola, Eric Gourgoulhon and Jean-Alain Marck}
\address{D\'epartement d'Astrophysique Relativiste et de Cosmologie \\
  UPR 176 du C.N.R.S., Observatoire de Paris, \\
  F-92195 Meudon Cedex, France}
\date{18 September 1997}
\maketitle

\begin{abstract}
A general relativistic version of the Euler equation for perfect fluid 
hydrodynamics is applied to a system of two neutron stars orbiting
each other. 
In the quasi-equilibrium phase of the evolution of this system, 
a first integral of motion can be derived for certain
velocity fields of the neutron star fluid including 
the (academic) case of co-rotation with
respect to the orbital motion (synchronized binaries) and the
realistic case of counter-rotation 
with respect to the orbital motion. 
The velocity field leading to this latter configuration can
be computed by solving three-dimensional vector and scalar Poisson equations.
\end{abstract}

\pacs{PACS number(s): 04.25.Dm, 04.30.Db, 04.40.Dg, 95.30.Sf, 97.60.Jd}

\section{Introduction} \label{s:intro}

Considerable efforts by many groups in the world 
are currently devoted to the
computation of the gravitational radiation from binary neutron star
coalescences (see e.g. \cite{Shapi96a} or \cite{Blanc96} for recent
reviews). These phenomena constitute  one of the most promising
sources of gravitational radiation for the interferometric detectors
GEO600, LIGO and VIRGO currently under construction 
\cite{CiufF96,MarcL96}. 
Basically two different approaches are used to tackle
this problem: 
\begin{description}
\item[A1] high-order Post-Newtonian analytical calculations
in the point-mass limit for the two coalescing objects
(see \cite{Blanc96} for a review);
\item[A2] hydrodynamical numerical
simulations which treat the neutron stars as perfect fluid 
balls. In this latter category, different
methods, based on different approximations, can be distinguished:
\begin{description}
\item[A2a] Newtonian 
affine approximations which consist in modeling the stars by 
triaxial ellipsoids, thereby reducing the dynamical degrees of freedom of
a star to a finite number and leading to ordinary differential equations
for the evolution, instead of partial differential equations
\cite{Kocha92,LaiRS94a,LaiRS94b,LaiSh95}; 
\item[A2b] Post-Newtonian affine approximations,
at the 1-PN order \cite{KokkS95,ShibT97,LomRS97}
or at the 2-PN order \cite{OgawK96};
\item[A2c] Newtonian hydrodynamical simulations, either based on 
finite difference methods \cite{NakaO91,ShiNO93,RufJS96,RuJTS97,RufRJ96} or on 
the  Smooth Particles Hydrodynamical (SPH) method 
\cite{ZhuCM94,ZhuCM96,RasiS94,DaBPT94}
(see ref.~\cite{RufRJ96} for a comparison of various codes);
\item[A2d] Post-Newtonian (at the order 1-PN) hydrodynamical simulations
with a finite difference method \cite{OohaN92,Shiba97};
\item[A2e] fully relativistic hydrodynamical simulations within the 3+1
formalism of general relativity and using the approximation of 
a conformally flat spatial 3-metric along with no gravitational field
dynamics \cite{WilsM95,WilMM96,BCSST97a,BSCST97,BCSST97b} 
(see also \cite{OohaN96}).
\end{description}
\end{description}

The analytical Post-Newtonian approach (A1) allows to compute the 
evolution of 
the binary system from an arbitrary early stage, when the separation
between the two components is large, up to the rapid inspiral phase
driven by the rapid loss of orbital energy by gravitational radiation. 
This approach breaks down when finite size effects (tidal
forces, disruption of one of the stars) become important, i.e. 
during the coalescence phase. 
This final stage can be studied only by means of the numerical hydrodynamical
methods (A2). But in this case one faces the problem of the initial conditions.
Indeed, due to the limitation of computer resources, the initial 
conditions cannot be set when the separation between the two stars
is much larger than their radii:
this would require a prohibitive number of time steps for the 
evolution codes: the time to coalescence increases with the fourth power of 
the initial separation $a_0$ between the two objects. 
In practice, all the fully hydrodynamical  
computations listed in (A2) have been performed with $a_0$ 
set to at most a few times the stellar radius $R$: $a_0 \simeq 5\, R$ in
ref.~\cite{DaBPT94}, 
$a_0 \simeq 4 \,R$
in refs.~\cite{ZhuCM94},\cite{ZhuCM96}, 
$a_0 \simeq 3\, R$ in refs.~\cite{ShiNO93},\cite{RufJS96},\cite{RufRJ96},
$a_0 = 2 \, R$ (!) in ref.~\cite{OohaN92}. 
For the calculations employing the affine approximation
instead of the full hydrodynamics, the initial separation is taken
to be somewhat larger: $a_0 \simeq 5 \, R$ in ref.~\cite{KokkS95},
$a_0 \simeq 15\, R$ in ref.~\cite{OgawK96}. 
Now at such small separations,
two effects are important: (i) 
tidal forces, i.e. the influence of the gravitational field of star 1 (resp. 
2) on the internal structure of star 2 (resp. 1),
and (ii) general relativity. 

Tidal effects have not been taken into
account in the computation of the initial conditions of most of the 
fully hydrodynamical studies
listed in (A2), the only exceptions being the works by 
Nakamura \& Oohara \cite{NakaO91}, Rasio \& Shapiro \cite{RasiS94} and
Baumgarte et al. \cite{BCSST97a,BSCST97,BCSST97b}, 
all in the case of {\em synchronized}
binaries, i.e. when the stars have zero spin in the frame co-moving with the
two centres of mass. 
However this rotation state is unrealistic. Indeed it can
be seen \cite{Kocha92,BildC92} that the 
neutron star matter shear viscosity is too small and the binary
evolution too rapid to lead to a
synchronization of the spin periods with the orbital period\footnote{As shown
by Kochanek \cite{Kocha92}, this
conclusion remains valid even if one takes into account the much higher
effective viscosity arising when the neutron star's solid crust enters 
the plastic flow regime.}. In other words, the
inspiral of the binary system can be described in terms of perfect fluids and, 
 in first approximation, all the forces acting on the fluid are
gradients of some scalar potentials (gravitational force,
gravitational radiation reaction force)\footnote{this is not true for
the so-called ``gravitomagnetic'' force; this latter induces
some circulation of the fluid, as studied recently by 
Shapiro \cite{Shapi96b}. 
However this effect is important only for neutron star - black hole binaries,
with a maximally rotating black hole.}, 
so that Kelvin's theorem applies:
the circulation of the fluid velocity (with respect to some inertial frame) 
on any closed contour comoving with
the star (e.g. the stellar equator) is conserved. For each star, 
the circulation around the equator is roughly $C\simeq 2 A \omega$ where
$A$ is the area of the equatorial cross section of the star and $\omega$
its mean angular velocity with respect to some inertial frame $\cal R$:
$\omega = \Omega_{\rm orb} + \Omega_{\rm spin}$ where $\Omega_{\rm orb}$
is the angular orbital velocity of the system with respect to $\cal R$ and
$\Omega_{\rm spin}$ is the rotation angular velocity of the star with
respect to the co-orbiting frame. Since the variation of $A$ is small
during the evolution to the coalescence, the conservation of $C$ is equivalent
to the conservation of $\omega$. When the separation is large $\Omega_{\rm orb}$
is negligible and $\omega$ is equal to the rotation rate of the
star. When the separation is that of the initial conditions of the 
numerical computations (A2) ($a \sim 50{\rm\, km}$) $\Omega_{\rm orb} \sim
2\times 10^3 {\rm\, rad\, s}^{-1}$, which is much larger than $\omega$, except
for neutron stars rotating initially at millisecond periods. 
Consequently, one must have 
\be \label{e:contre_rot}
\Omega_{\rm spin} = - \Omega_{\rm orb} \ .
\ee
in order to have a constant circulation. We call configurations 
obeying (\ref{e:contre_rot}) {\em counter-rotating configurations}. They
represent realistic initial conditions for neutron star binary coalescence. 

Some of the fully hydrodynamical calculations listed in (A2) employ
(\ref{e:contre_rot}) as initial conditions 
\cite{ShiNO93,RufJS96,DaBPT94}. 
But none of them take into account the tidal effects:
the stars are taken to be either spherical \cite{RufJS96,DaBPT94}
or axisymmetric \cite{ShiNO93} (as mentioned above, the only computations 
with self-consistent initial conditions concerns synchronized binaries
\cite{NakaO91,RasiS94,BCSST97a,BSCST97,BCSST97b}, 
for which $\Omega_{\rm spin}=0$).
To date, the only self-consistent initial conditions obeying 
(\ref{e:contre_rot})
have been computed by  Bonazzola \& Marck \cite{BoGHM92}. 
As can be seen in Fig.~3 
of ref.~\cite{BoGHM92}, the tidal deformation is quite important when
the separation is $a_0 \simeq 3\, R$.
However, these initial conditions have not been employed
in evolution calculations. 

It must be noticed that some of the studies performed in the affine
approximation (A2a) make use of self-consistent counter-rotating initial
conditions. They correspond to {\em irrotational Darwin-Riemann
ellipsoids} \cite{LaiRS94a,LaiSh95} or 
(in the approximation of a large
separation) {\em irrotational Roche-Riemann ellipsoids} \cite{Kocha92}.  

As regards general relativistic effects,
the often used Newtonian approximation [items (A2a) and 
(A2c)] is very crude, in particular for the neutron star internal 
structure: let us recall that for a typical $1.4\, M_\odot$ neutron
star, the central value of the metric coefficient $g_{00}$ is around 
$0.4$, which shows that even a (first order) Post-Newtonian
approximation is not sufficient for describing these objects.

The purpose of the present article is to give a method for
computing self-consistent (i.e.
including the tidal and rotational distortion) and realistic
(i.e. obeying (\ref{e:contre_rot})) initial conditions for
binary neutron stars in the framework of the full general
relativity. Therefore, this work can be conceived as the
extension to general relativity of Bonazzola \& Marck results \cite{BoGHM92}.

The envisaged problem can be decomposed in two parts: (i) the computation
of the gravitational field (i.e. the spacetime metric) generated by
the two stars and their motion and (ii) the computation of the stellar
structure (density distribution, velocity field,...) in that 
gravitational field. Part (i) is the main topic of numerical relativity
and can be, at least in principle, be achieved by means of the classical
3+1 formalism (see e.g. \cite{SmarY78} and \cite{Abrah96}).  
This paper focuses on the determination of the stellar structure. 
For this purpose,
we consider that in the vicinity of the searched initial conditions,
the system 
evolves along a sequence of equilibrium states. We obtain a first
integral of motion for certain classes of velocity field inside
the neutron stars, including the  
co-rotating and the (realistic) counter-rotating cases.

The plan of the paper is as follows. 
Sect.~\ref{s:sym} translates the basic assumption of quasi-equilibrium
in geometrical terms (a spacetime symmetry) 
which leads to the definition of a privileged
observer (the ``co-orbiting'' observer). The (relativistic)
Euler equation for the fluid velocity is then derived in the frame
of that observer (Sect.~\ref{s:Euler}). 
Necessary and sufficient conditions to get a first
integral of the Euler equation are given in Sect.~\ref{s:first_int}.
This first integral is trivially obtained in the (unrealistic) case
of co-rotating stars. The astrophysically relevant case of 
counter-rotating stars is presented in 
Sect.~\ref{s:cont-rot}. A method of resolution is discussed in 
Sect.~\ref{s:discuss}.

\section{Spacetime symmetry and choice of coordinates} 
\label{s:sym}

\subsection{Quasi-equilibrium hypothesis} \label{s:quasi-equil}

\begin{figure}
\centerline{ \epsfig{figure=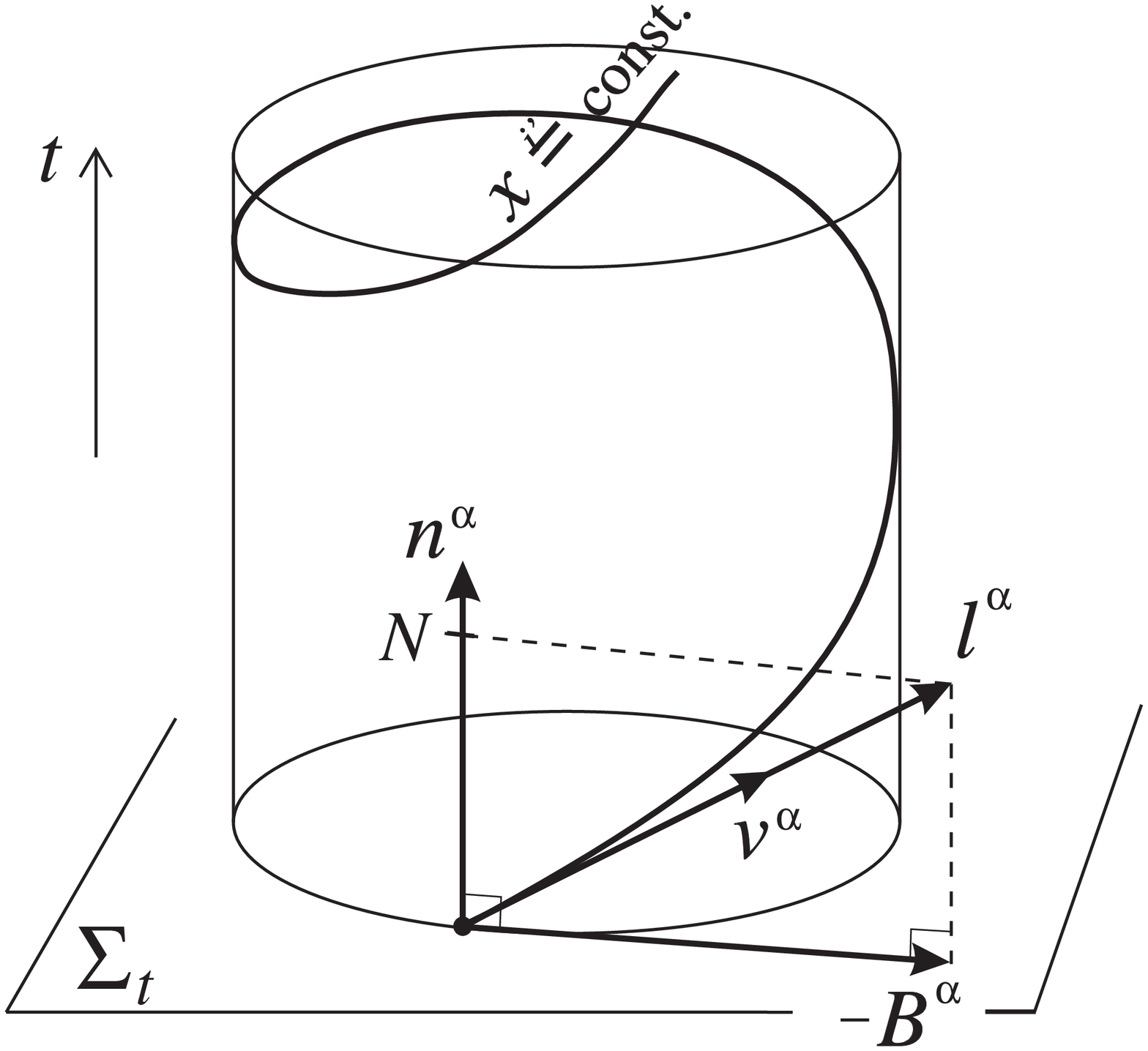,height=8cm} }
\vspace{1cm}
\caption[]{\label{f:heli}   
Spacetime foliation $\Sigma_t$, helicoidal Killing vector
$l^\alpha$ and its trajectories $x^{i'} = {\rm const}$, which are the 
worldlines of the co-orbiting observer (4-velocity: $v^\alpha$).
Also shown are the rotating-coordinate shift vector $B^\alpha$
and the unit future-directed vector $n^\alpha$, normal to the spacelike
hypersurface $\Sigma_t$.} 
\end{figure}

When the separation between the centres of the two neutron stars
is about $50{\rm\, km}$ (in harmonic coordinates)
the time variation of the orbital period, $\dot P_{\rm orb}$, computed at 
the 2-PN order by means of the formulas established by 
Blanchet et al. \cite{BDIWW95} is about $2\%$.
The evolution at this stage can thus be still considered as a sequence
of equilibrium configurations.
Moreover the orbits are expected to be circular (vanishing eccentricity), 
as a consequence of the gravitational radiation reaction \cite{Peter64}.
In terms of the spacetime geometry, 
we translate these assumptions by
demanding that there exists a Killing vector field $l^\alpha$ which is
expressible as
\be \label{e:helicoidal}
 l^\alpha = k^\alpha + \Omega_{\rm orb} \, m^\alpha \ ,
\ee
where $\Omega_{\rm orb}$ is a constant, to be identified with the orbital
angular velocity with respect to a distant inertial observer, and
$k^\alpha$ and $m^\alpha$ are two vector fields with the following 
properties. $k^\alpha$ is timelike at least far from the binary and 
is normalized so that far from the star it coincides
with the 4-velocity of inertial observers. $m^\alpha$ is spacelike,
has closed orbits, is zero on a two dimensional timelike surface, called
the {\em rotation axis}, and is normalized so that 
$\nabla_\mu(m_\rho m^\rho) \nabla^\mu(m_\sigma m^\sigma) / (4 m_\nu m^\nu)$
tends to $1$ on the rotation axis [this latter condition ensures that
the parameter $\phi$ associated with $m^\alpha$ along its trajectories
by $m^\alpha = (\partial/\partial \phi)^\alpha$ has the standard $2\pi$
periodicity].
Let us call $l^\alpha$ the {\em helicoidal Killing vector}. We assume
that $l^\alpha$ is a symmetry generator not only for the spacetime
metric $g_{\alpha\beta}$ but also for all the matter fields. In particular,
$l^\alpha$ is tangent to the world tubes representing the surface of 
each star, hence its qualification of {\em helicoidal} 
(cf. Figure~\ref{f:heli}).

The approximation suggested above amounts to neglect outgoing gravitational
radiation. For non-axisymmetric systems --- as binaries are ---
it is well known that imposing $l^\alpha$ as an exact 
Killing vector leads to a spacetime which is not asymptotically flat
\cite{BlacD92}. Thus, in solving for the gravitational field equations,
a certain approximation has to be devised in order to 
avoid the divergence of some metric coefficients at infinity. For instance
such an approximation could be the Wilson \& Mathews scheme \cite{WilsM89}
that amounts to
solve only for the Hamiltonian and momentum constraint equations.
This approximation has been used in all the fully relativistic
studies to date \cite{WilsM95,WilMM96,BCSST97a,BSCST97,BCSST97b} 
and is consistent 
with the existence of the helicoidal Killing vector field (\ref{e:helicoidal}).
Note also that since the gravitational radiation reaction shows up
only at the 2.5-PN order, the helicoidal symmetry is exact up to the
2-PN order.

\subsection{3+1 Foliation of spacetime} \label{s:3+1fol}

For the considered problem, two types of coordinates can be envisaged:
``non-rotating'' coordinates $(t,x^i)$
which are Minkowskian at infinity,
so that $k^\alpha$ is the first vector of the natural basis
corresponding to these coordinates and ``co-rotating'' coordinates 
$(t',x^{i'})$ so
that $l^\alpha$ is the first vector of their natural basis. 
There is a lot of ways to do this. We choose both coordinate systems
so that the hypersurfaces $t={\rm const}$ and $t'={\rm const}$ coincides
and are maximal spacelike hypersurfaces. More precisely, we suppose that
there exists a slicing of spacetime by a family of spacelike hypersurfaces
$(\Sigma_t)$ so that (i) each $\Sigma_t$ is spacelike and (ii) $m^\alpha$ is
tangent to $\Sigma_t$. 

\subsubsection{Non-rotating coordinates}

On each $\Sigma_t$, we choose a system of Cartesian coordinates
$x^i = (x,y,z)$, such that $(t,x^i)$ is a system of spacetime coordinates
satisfying to 
\begin{eqnarray}
	k^\alpha & = & \l( {\partial\ov \partial t } \r) ^\alpha  \\
	m^\alpha & = & - y \l( {\partial\ov \partial x } \r) ^\alpha 
	+ x \l( {\partial\ov \partial y } \r) ^\alpha \ .
\end{eqnarray}
The {\em lapse function} $N$ and {\em shift vector} $N^\alpha$ 
associated with these coordinates are defined by 
\be \label{e:k,lapse,shift}
	k^\alpha = N n^\alpha - N^\alpha  \qquad \mbox{and}
				\qquad n_\mu N^\mu = 0 \ ,
\ee
where $n^\alpha$ is the future directed unit 4-vector normal to the 
hypersurface $\Sigma_t$.

\subsubsection{Rotating coordinates} \label{s:rot_coord}

We call {\em rotating coordinates} any coordinate system 
$x^{i'}$ on $\Sigma_t$ such that $(t'=t,x^{i'})$ is a spacetime coordinate
system satisfying to 
\be
	l^\alpha =  \l( {\partial\ov \partial t' } \r) ^\alpha \ .
\ee
In other words, the lines $x^{i'} = {\rm const}$ are the 
trajectories of $l^\alpha$. This latter being a Killing vector, this means
that $t'$ is an {\em ignorable} coordinate for such systems. 
In numerical studies, we will use these types of coordinates to reduce the
a priori 4-D problem to a 3-D one. In practice, three 
rotating coordinate systems can be used: one centered on each star, to 
describe properly the hydrodynamics,  
and a third one centered on the rotation axis, to describe the gravitational
field. 
The lapse function and shift vector associated with rotating coordinates
are immediately deduced from Eqs.~(\ref{e:helicoidal}) and 
(\ref{e:k,lapse,shift}) which result in
\be \label{e:l=Nn-B}
   l^\alpha = N n^\alpha - B^\alpha 	\ ,
\ee
with
\be \label{e:B:def}
	B^\alpha  := N^\alpha - \Omega_{\rm orb}\, m^\alpha \ .
\ee
Since $m^\alpha$ is parallel to $\Sigma_t$, $B^\alpha$ is indeed the
shift associated with rotating coordinates (cf. Figure~\ref{f:heli}).
Note that rotating and
non-rotating coordinates have the same lapse $N$ for they define 
the same spacetime foliation. 

The Killing equation $\nabla_\alpha l_\beta + \nabla_\beta l_\alpha = 0$,
once projected onto $\Sigma_t$, leads to the following relation between
the extrinsic curvature tensor $K_{\alpha\beta}$ of the 
hypersurfaces $\Sigma_t$ and the derivatives of the shift vector
$B^\alpha$:
\be
   2 N K_{\alpha\beta} = - \nabla_\alpha B_\beta - \nabla_\beta B_\alpha
	- 2 n_\alpha n_\beta n^\mu \nabla_\mu N 
\ee
or, equivalently,
\be \label{e:K=DB+DB}
   2 N K_{\alpha\beta} = - \hat\nabla_\alpha B_\beta 
		- \hat\nabla_\beta B_\alpha \ ,
\ee
where $\hat\nabla_\alpha$ stands for the covariant derivative associated
with the 3-metric $h_{\alpha\beta}$
induced by $g_{\alpha\beta}$ onto the hypersurfaces $\Sigma_t$.
Note that $K_{\alpha\beta}$ is linked to the covariant derivative of
$n^\alpha$ by the formula
\be
	\nabla_\beta n_\alpha = - K_{\alpha\beta} 
		- \hat\nabla_\alpha(\ln N) \, n_\beta \ .
\ee
In the following an extensive use is made of this relation, without
explicitly mention it.

\subsection{The co-orbiting observer}

Let us call the {\em co-orbiting observer} the
observer $\cal O$ whose world lines coincide with the
trajectories of the symmetry group when these latter are timelike, which 
encompasses the region occupied by the two stars. $\cal O$'s  4-velocity 
$v^\alpha$ can be written:
\be \label{e:v=exp_l}
    v^\alpha = e^{-\Phi} \, l^\alpha \ ,
\ee
where $\Phi$ is a scalar field which is uniquely specified by the 
normalization
relation $v_\mu v^\mu = -1$.  It can be seen easily that $\Phi$ is 
related to the lapse $N$, the shift vectors $B^\alpha$ and $N^\alpha$,
the azimuthal vector $m^\alpha$ and the orbital angular velocity 
$\Omega_{\rm orb}$ by
\be \label{e:exp(2Phi)}
  e^{2 \Phi} = N^2 - B_\mu B^\mu = N^2 - \Omega_{\rm orb}^2 m_\mu m^\mu 
	+ 2 \Omega_{\rm orb} m_\mu N^\mu - N_\mu N^\mu \ . 
\ee

Let $q_{\alpha\beta}$ be the projector onto the 3-planes $\Pi$ 
orthogonal to $v^\alpha$: 
\be
    q_{\alpha\beta} := g_{\alpha\beta} + v_\alpha\, v_\beta \ . 
\ee
The kinematics of the observer $\cal O$ is entirely specified by 
the Ehlers decomposition \cite{Ehler61} of the covariant (with 
respect to $g_{\alpha\beta}$) derivative of $v^\alpha$:
\be
   \nabla_\beta v_\alpha = \omega_{\alpha\beta} + \theta_{\alpha\beta}
	- a_\alpha v_\beta \ ,
\ee
where
\be
   \omega_{\alpha\beta} := q_\alpha^{\ \, \mu} q_\beta^{\ \, \nu}
	\nabla_{[\nu} \, v_{\mu]}
\ee
is the {\em rotation 2-form} of $\cal O$,
\be \label{e:expansion,tens}
  \theta_{\alpha\beta} := q_\alpha^{\ \, \mu} q_\beta^{\ \, \nu}
	\nabla_{(\nu} \, v_{\mu)} 
\ee
is the {\em expansion tensor} of $\cal O$ and
\be
  a_\alpha := v^\mu \nabla_\mu v_\alpha 
\ee
is the {\em 4-acceleration} of $\cal O$. 
The property (\ref{e:v=exp_l}), namely that $v^\alpha$ is collinear to a 
Killing vector, means that $v^\alpha$ is an {\em isometric flow} \cite{Ehler61}
and leads to 
\be \label{e:theta=0}
    \theta_{\alpha\beta} = 0
\ee
and
\be \label{e:a=grad_phi} 
   a_\alpha =  \nabla_\alpha \Phi \ . 
\ee
Equation (\ref{e:theta=0}) shows that $v^\alpha$ is a rigid flow.

The three-dimensional vector space $\Pi$ represents the local rest frame
of $\cal O$. Note that since $\cal O$ is rotating 
($\omega_{\alpha\beta}\not = 0$, see below), $\Pi$ is not integrable into
global 3-surfaces. 
$q_{\alpha\beta}$ is the (positive definite) metric tensor induced by 
$g_{\alpha\beta}$ on $\Pi$. 
We can introduce  the alternating tensor within $\Pi$ as
\be \label{e:bareps,def}
  \bar\epsilon_{\alpha\beta\gamma} := v^\mu \epsilon_{\mu\alpha\beta\gamma} \ ,
\ee
where $\epsilon_{\alpha\beta\gamma\delta}$ is the spacetime alternating
tensor associated with the spacetime metric $g_{\alpha\beta}$. 
The rotation 2-form of $\cal O$ is fully specified by its dual within $\Pi$:
$\omega_{\alpha\beta} = - \omega^\mu \bar\epsilon_{\mu\alpha\beta}$, where
\be \label{e:omega_def}
  \omega^\alpha := - {1\ov 2} \bar\epsilon^{\,\alpha\mu\nu} \omega_{\mu\nu}
	= {1\ov 2} \bar\epsilon^{\,\alpha\mu\nu}  \nabla_\mu v_\nu \ .
\ee
Note that the Raychaudhuri identity for the flow $v^\alpha$ reduces
to a simple relation between the norm of $\omega^\alpha$ and the
Laplacian of $\Phi$:
\be
   \omega_\mu \omega^\mu = - \nabla_\mu \nabla^\mu \Phi 
		+ {1\ov 2} R_{\mu\nu} v^\mu v^\nu \ , 
\ee
where $R_{\alpha\beta}$ is the Ricci tensor of the metric $g_{\alpha\beta}$. 

By means of Eqs.~(\ref{e:v=exp_l}), (\ref{e:l=Nn-B}), (\ref{e:B:def})
and the Killing identity 
$\nabla_\alpha l_\beta + \nabla_\beta l_\alpha = 0$, the rotation vector
(\ref{e:omega_def}) can be expressed as
\be \label{e:omega=Omega}
  \omega^\alpha = {e^{-\Phi}\ov 2} \, \bar\epsilon^{\,\alpha\mu\nu} \l[
	\Omega_{\rm orb}  \nabla_\mu m_\nu - \nabla_\mu N_\nu  
  + 2 ( \Omega_{\rm orb}  m_\mu - N_\mu) \nabla_\nu \ln N  \r] \ .
\ee

\section{Relativistic Euler equation in the rotating frame} 
\label{s:Euler}

\subsection{Fluid motion}

As stated in the introduction, the matter constituting the neutron stars
can be considered as a perfect fluid, so that its stress-energy tensor
writes
\be 
    T^{\alpha\beta} = (e+p) \, u^\alpha u^\beta + p\, g^{\alpha\beta} \ .
\ee
The fluid 4-velocity $u^\alpha$ can be decomposed orthogonally 
with respect to the rotating observer $\cal O$ as follows:
\be \label{e:decomp_u}
   u^\alpha  =  \Gamma (V^\alpha + v^\alpha)  \ , 
\ee
where $\Gamma$ is the Lorentz factor
\be
    \Gamma := - v_\mu u^\mu \ ,
\ee
and $V^\alpha$ is the fluid {\em 3-velocity} with respect to $\cal O$:
\be
    V^\alpha := {1\ov \Gamma} q^\alpha_{\ \, \mu} u^\mu \ .
\ee
$V^\alpha$ belongs to $\Pi$ and 
is the fluid velocity as measured by the observer $\cal O$ (i.e. with 
respect to $\cal O$'s proper time). 
As an immediate consequence of $u_\mu u^\mu  = -1$,
one has the usual relation between $\Gamma$ and $V^\alpha$:
\be \label{e:Lorentz}
   \Gamma  =  (1-V_\mu V^\mu)^{-1/2} \ .
\ee

To a very good approximation the (cold)
neutron star matter equation of state (hereafter EOS)
can be considered as barotropic: $e=e(n)$ and
$p=p(n)$ where $n$ is the proper baryon density. It is then worth to
introduce the logarithm of the ratio of enthalpy per baryon and the baryon 
mean rest mass $m_B$ by
\be
   H := \ln \l( {e + p\ov m_{\rm B} n} \r)  \ , 
\ee
which we call the {\em log-enthalpy},
to re-express $(\nabla_\alpha p)/(e+p)$
as a gradient of a scalar: indeed, by virtue of the First Law of 
Thermodynamics, the following identity holds for any barotropic EOS:
\be
    {\nabla_\alpha p \ov e+p } = \nabla_\alpha H \ .
\ee

Then the fundamental energy-momentum conservation equation 
\be \label{e:divT=0}
   \nabla_\mu T^{\mu\alpha} = 0 
\ee
can easily be shown to be equivalent to the system of the following
two equations
\begin{eqnarray}
    & & \nabla_\mu ( n u^\mu) = 0 		\label{e:div(nu)=0} \\
    & & u^\mu \nabla_\mu u^\alpha + \nabla^\alpha H 
	+ u^\mu \nabla_\mu H \, u^\alpha = 0  \label{e:mouv,canon} \ . 
\end{eqnarray}
Note that Eq.~(\ref{e:mouv,canon}) is nothing but the
{\em uniformly canonical equation of motion} for a single-constituent perfect 
fluid as given by Carter \cite{Carte79}.

\subsection{Baryon number conservation}

Inserting Eq.~(\ref{e:decomp_u}) into the baryon number conservation
equation (\ref{e:div(nu)=0}) and using the fact that $v^\alpha$ is
divergence-free [Eq.~(\ref{e:theta=0})] leads to
\be \label{e:baryon,cons}
     \nabla_\mu V^\mu + V^\mu \nabla_\mu \ln (n\Gamma) = 0 \ . 
\ee

\subsection{Momentum conservation}

By projecting Eq.~(\ref{e:mouv,canon}) onto $v^\alpha$ (i.e. along
the world lines of the co-orbiting observer $\cal O$), one obtains
the relation
\be \label{e:Bernouilli,V}
   V^\mu \nabla_\mu \l( H + \Phi + \ln\Gamma \r) = 0 \ ,
\ee
which can be considered as a
relativistic generalization of the classical Bernouilli
theorem, for it means that the quantity $H + \Phi + \ln\Gamma$ is constant 
along the fluid lines.

By projecting Eq.~(\ref{e:mouv,canon}) perpendicularly to $v^\alpha$ 
(i.e. onto the local rest frame of the co-orbiting observer $\cal O$), 
one obtains the relativistic version of the Euler equation for the
fluid velocity with respect to $\cal O$ :
\be \label{e:Euler,p}
 V^\mu \nabla_\mu V^\alpha +2 \bar\epsilon^{\, \alpha}_{\ \ \mu\nu}
	\omega^\mu V^\nu + \nabla^\alpha \Phi 
	- V^\mu \nabla_\mu \Phi \ V^\alpha + \Gamma^{-2} \nabla^\alpha H
	= 0 \ .
\ee
At the Newtonian limit, the $V^\mu \nabla_\mu V^\alpha$
gives the classical term $(\vec{V}\cdot\vec{\nabla})\vec{V}$
and $2 \bar\epsilon^{\, \alpha}_{\ \ \mu\nu} \omega^\mu V^\nu$ 
gives the Coriolis term
$2\, \vec\omega\times\vec V$, induced by the
rotation of the observer $\cal O$ with respect to some inertial frame.   
The term $\nabla^\alpha H$ gives the 
classical pressure term. Finally $\Phi$
reduces to the sum of the gravitational and centrifugal potentials
[cf. Eq.~(\ref{e:exp(2Phi)})]
\be \label{e:Phi_Newt}
   \Phi \stackrel{{\rm Newt.}}{\longrightarrow} \Phi_{\rm grav} 
   - {1\ov 2}     \l( \vec\Omega_{\rm orb} \times \vec{r} \r) ^2 \ ,
\ee
$\Phi_{\rm grav}$ being defined so that the gravitational field
writes $\vec{g} = - \vec\nabla \Phi_{\rm grav}$.

In order to exhibit from Eq.~(\ref{e:Euler,p}) a first integral of  
motion, we shall write as much
terms as possible under the form of gradients. First, it can be seen
easily that, similarly to the usual flat space formula, the following
identity holds
\be \label{e:V.gradV}
   V^\mu \nabla_\mu V^\alpha = 
	\bar\epsilon^{\, \alpha}_{\ \ \mu\nu} (\nabla\wedge V)^\mu
	V^\nu + \nabla^\alpha \l( {1\ov 2} V_\mu V^\mu \r) \ , 
\ee
where we have introduced the curl of $V^\alpha$ within the 3-space
$\Pi$:
\be \label{e:curl:def}
    (\nabla\wedge V)^\alpha := 
  \bar\epsilon^{\, \alpha}_{\ \ \mu\nu} \nabla^\mu V^\nu \ .
\ee

Putting Eq.~(\ref{e:V.gradV}) into Eq.~(\ref{e:Euler,p}) and
performing slight rearrangements results in
\be \label{e:Euler,grad}
   \nabla_\alpha ( H + \Phi + \ln\Gamma )
  + \Gamma^2 \l\{ \bar\epsilon_{\alpha\mu\nu} 
   \l[ (\nabla \wedge V)^\mu + 2 \omega^\mu \r] V^\nu
  + V_\mu V^\mu P_\alpha^{\, \ \nu} \nabla_\nu \Phi \r\} 
	= 0 \ ,
\ee
where 
\be
    P_\alpha^{\,\ \beta} := q_\alpha^{\,\ \beta} - {V_\alpha V^\beta \ov
       V_\mu V^\mu } 
\ee
is the projector onto the 2-space orthogonal to $V^\alpha$, i.e. 
orthogonal to the fluid lines
with respect to $\cal O$. Note that in the case where the fluid is at
rest with respect to $\cal O$, $P_\alpha^{\,\ \beta}$ is not defined;
however, the product $V_\mu V^\mu P_\alpha^{\, \ \beta}$ which appears in
Eq.~(\ref{e:Euler,grad}) remains well defined and is equal to zero. 
In the derivation
of Eq.~(\ref{e:Euler,grad}), use has been made of Eq.~(\ref{e:Lorentz})
to replace the term $\nabla_\alpha(V_\mu V^\mu /2)$ coming from
Eq.~(\ref{e:V.gradV}) by $\Gamma^{-2} \nabla_\alpha \ln\Gamma$.

\subsection{Number of independent components} \label{s:n_indep}

From the fundamental equation $\nabla_\mu T^{\mu\alpha} = 0$, which
has a priori four independent components, we have derived 
two scalar equations [Eqs.~(\ref{e:baryon,cons}) and
Eqs.~(\ref{e:Bernouilli,V})] and one vectorial equation
[Eq.~(\ref{e:Euler,grad})]. 
Equations~(\ref{e:Bernouilli,V}) and (\ref{e:Euler,grad}) are not 
independent: the former is a direct consequence of the latter, as 
seen easily by projecting (\ref{e:Euler,grad}) along $V^\alpha$.
Moreover, from its construction, (\ref{e:Euler,grad}) has only 
three independent components for it lies into the 3-planes $\Pi$ orthogonal
to $v^\alpha$. 

We will take the scalar equation (\ref{e:baryon,cons}) and the
vectorial equation lying in $\Pi$ (\ref{e:Euler,grad}) as the
fundamental equations to be satisfied for our problem.

\section{Constraint on the velocity field and first integral of motion}
\label{s:first_int}

The only assumption underlying Eq.~(\ref{e:Euler,grad}) is that
the observer $\cal O$, with respect to which the fluid velocity 
$V^\alpha$ is defined, has world lines parallel to the helicoidal 
Killing vector $l^\alpha$. Equation~(\ref{e:Euler,grad}) is
equivalent to the system
\be \label{e:cond,G}
  \bar\epsilon_{\alpha\mu\nu} 
   \l[ (\nabla \wedge V)^\mu + 2 \omega^\mu \r] V^\nu
  + V_\mu V^\mu P_\alpha^{\, \ \nu} \nabla_\nu \Phi 
	= \Gamma^{-2} \nabla_\alpha G 
\ee
\be \label{e:intpre}
    H + \Phi + \ln\Gamma + G = {\rm const.} \ ,
\ee
where $G$ is a scalar field defined at least in the stars' world tubes.
The relation (\ref{e:intpre}) constitutes a first integral of motion
of the system.

\subsection{The co-rotating case}

Equation (\ref{e:cond,G}) is trivially satisfied in the case 
$V^\alpha = 0$ by taking $G={\rm const}$. The first integral of 
motion reduces then to
\be \label{e:intpre_synchr}
    H + \Phi = {\rm const.}
\ee
This case is the co-rotating one ($\Omega_{\rm spin}=0$) mentioned in the
introduction; it corresponds to {\em synchronized} binaries, which are
not expected to represent realistic close neutron star binaries,
as discussed in Sect.~\ref{s:intro}.

The first integral (\ref{e:intpre_synchr}) follows directly from the
fact that in the co-rotating case the fluid 4-velocity is parallel to a 
Killing vector \cite{BonFG96}:
$V^\alpha=0$ is indeed equivalent to 
$u^\alpha=v^\alpha = e^{-\Phi} l^\alpha$
[cf. Eq.~(\ref{e:decomp_u})]. The integral (\ref{e:intpre_synchr})
 is well known in the case of a single
rigidly rotating star (see e.g. \cite{BoGSM93}). For the problem of 
the initial conditions of a binary coalescence, it has been used 
 by Nakamura \& Oohara \cite{NakaO91} (at the Newtonian approximation)
and Baumgarte et al. \cite{BCSST97a,BSCST97,BCSST97b}.

\subsection{Formulation of the problem in the general case}

From now on, we suppose that $V^\alpha \not = 0$. By performing
the vector product (with respect to $\bar\epsilon_{\alpha\beta\gamma}$)
of Eq.~(\ref{e:cond,G}) by $V^\alpha$, one can see easily that 
Eq.~(\ref{e:cond,G}) is equivalent to the system
\begin{eqnarray}
  & & V^\mu \nabla_\mu G =  0 \label{e:VgragG=0} \\
 & &  (\nabla\wedge V)^\alpha =  - 2\omega^\alpha
	- \bar\epsilon^{\, \alpha\mu\nu} V_\mu \nabla_\nu \Phi
        + (\Gamma^2 V_\sigma V^\sigma)^{-1} 
     \bar\epsilon^{\, \alpha\mu\nu} V_\mu \nabla_\nu G
     + F \, V^\alpha \ , \label{e:rotV,F,G}
\end{eqnarray}
where 
\be \label{e:F:def} 
   F := {1\ov V_\nu V^\nu} V_\mu
    \l[ (\nabla\wedge V)^\mu + 2\omega^\mu \r] \ .
\ee

The gravitational field being given, 
the problem of getting a solution amounts to finding
a vector field $V^\alpha$ and a scalar field $G$ such
that the equations (\ref{e:baryon,cons}), (\ref{e:VgragG=0}) and
(\ref{e:rotV,F,G}) are satisfied. In Eq.~(\ref{e:baryon,cons}), the
scalar field $n$ is that related to the gravitational field, $V$
and $G$ by the first integral (\ref{e:intpre}) via the EOS
$n=n(H)$.
More precisely, let us consider an iterative method for solving this
problem. Let us suppose that at a given step, the gravitational
field equations have been solved; the potential $\Phi$ and the
rotation vector $\omega^\alpha$ are then known. The enthalpy $H$
can be then deduced from the first integral (\ref{e:intpre}), by
taking for $\Gamma$ and $G$ the values at the previous step or making
some extrapolation from a few previous steps. The baryon density $n$
is computed from $H$ by means of the EOS.  
The system of 
equations (\ref{e:baryon,cons}), (\ref{e:VgragG=0}) and
(\ref{e:rotV,F,G}) is then to be solved in $V^\alpha$ and $G$.
It is however not obvious that a solution  exists in the general case.
What can be said is that in the Newtonian and incompressible case,
solutions do exist and are constituted by S-type Darwin-Riemann 
ellipsoids \cite{Aizen68}.

\section{The counter-rotating case}
\label{s:cont-rot}

\subsection{Definition}

Let us focus on the interesting case of counter-rotating binaries.
The concept of counter-rotation has to be defined in the relativistic
framework. We shall define it by requiring $V^\alpha \not = 0$ and
the scalar field $G$ introduced in Eqs.~(\ref{e:cond,G})-(\ref{e:intpre})
to be constant:
\be \label{e:G=const}
    G = {\rm const.}
\ee
This definition is motivated by the fact that at the Newtonian 
limit\footnote{Details about the Newtonian limit will be presented 
in Sect.~\ref{s:Newt_lim}.}
it implies $1/2 \, (\nabla\wedge V)^\alpha =  - \omega^\alpha$, which
is the definition (\ref{e:contre_rot}) of counter-rotation
[cf Eq.~(\ref{e:omega=Omega})].

With the choice (\ref{e:G=const}), Eq.~(\ref{e:VgragG=0}) is trivially 
satisfied and Eq.~(\ref{e:rotV,F,G}) becomes
\be \label{e:rotV,F}
(\nabla\wedge V)^\alpha =  - 2\omega^\alpha
	- \bar\epsilon^{\, \alpha\mu\nu} V_\mu \nabla_\nu \Phi
     + F \, V^\alpha \ . 
\ee

\subsection{3+1 decomposition}

From the numerical point of view, it is desirable to reduce the 
problem to the resolution of three-dimensional equations.
Now Eq.~(\ref{e:rotV,F}) involves 4-vectors: even if $V^\alpha$ is
spacelike and belongs to 3-planes orthogonal to $v^\alpha$,
due to the rotation of this latter, there exists
no coordinate system in which $V^\alpha$ would have only three 
non-vanishing components. Therefore, we choose to 
recast Eq.~(\ref{e:rotV,F}) according to the 3+1 foliation of spacetime 
introduced in Sect.~\ref{s:3+1fol}. In this manner, we will consider 
only 3-vectors belonging to the spacelike hypersurfaces $\Sigma_t$.
The first step is to introduce the orthogonal decomposition with respect
to $\Sigma_t$ of the fluid velocity $V^\alpha$ with respect to the
co-orbiting observer:
\be \label{e:V=W+Z}
	V^\alpha = W^\alpha + Z n^\alpha \ ,
\ee
where
\be
  W^\alpha = h^\alpha_{\ \, \mu} V^\mu \qquad
	\mbox{and} \qquad 
  Z = -n_\mu V^\mu \ ,
\ee
where $h_{\alpha\beta} := g_{\alpha\beta} + n_\alpha \, n_\beta$ is
the orthogonal projector onto $\Sigma_t$, or equivalently, the
3-metric induced by $g_{\alpha\beta}$ in $\Sigma_t$.
Due to the orthogonality relation $v_\mu V^\mu = 0$,  
the scalar $Z$ is not independent from $W^\alpha$:
by inserting Eqs.~(\ref{e:v=exp_l}) and (\ref{e:l=Nn-B}) into 
$v_\mu V^\mu = 0$, one gets
\be \label{e:Z(B,W)}
  Z = - {1\ov N} B_\mu W^\mu = - {1\ov N} B_i W^i \ .
\ee
In the last part of this equation, we have introduced Latin indices,
which range from 1 to 3, whereas the Greek indices range from 0 to 3.
We will systematically do this in the following for all the tensor fields
that lie in $\Sigma_t$, such the vectors $B^\alpha$ and $W^\alpha$.
In this way the three-dimensional character of the equations will 
clearly appear.

The curl on the left-hand side of Eq.~(\ref{e:rotV,F}) is defined within
respect to the alternating tensor $\bar\epsilon^{\alpha\beta\gamma}$,
which is neither parallel nor orthogonal to the $(\Sigma_t)$ foliation.
Let us introduce
instead the alternating tensor $\hat\epsilon^{\alpha\beta\gamma}$
within the space $(\Sigma_t,\, h_{\alpha\beta})$ by
\be
  \hat\epsilon^{\alpha\beta\gamma} := 
	n_\mu \epsilon^{\mu\alpha\beta\gamma} \ .
\ee
Inserting the identity
$\epsilon^{\alpha\beta\gamma\delta} = -4 n^{[\alpha} 
	\hat\epsilon^{\beta\gamma\delta]} $
into the definition (\ref{e:bareps,def}) of 
$\bar\epsilon^{\alpha\beta\gamma}$, we arrive at the following
expression of $\bar\epsilon^{\alpha\beta\gamma}$ in term of
$\hat\epsilon^{\alpha\beta\gamma}$:
\be \label{e:bar_eps=...}
  \bar\epsilon^{\alpha\beta\gamma} = e^{-\Phi} (
	N \hat\epsilon^{\alpha\beta\gamma} 
	- n^\alpha B_\mu \hat\epsilon^{\mu\beta\gamma}
	- n^\beta  B_\mu \hat\epsilon^{\mu\gamma\alpha}
	- n^\gamma  B_\mu \hat\epsilon^{\mu\alpha\beta} ) \ .
\ee

Besides, we can express the four-dimensional
covariant derivative of $V^\alpha$ which appears in the
curl of $V^\alpha$
in terms of the three-dimensional covariant derivative of $W^\alpha$
with respect to $h_{\alpha\beta}$ :
\be \label{e:DV=...}
  \nabla_\alpha V_\beta = \hat\nabla_\alpha W_\beta 
	- n_\alpha n^\mu \nabla_\mu W_\beta
	- K_{\alpha\mu} W^\mu n_\beta
	+ n_\beta \nabla_\alpha Z - Z K_{\alpha\beta}
	- Z n_\alpha \hat\nabla_\beta\ln N \ .
\ee

A useful formula is that which gives the derivative along $n^\alpha$ of
any tensor field $X^\alpha$ (i) which lies in $\Sigma_t$ and (ii) which respects 
the heloicoidal symmetry (i.e. ${\cal L}_{\bf l} X^\alpha = 0$):
\be \label{e:n.grad(X)}
   n^\mu \nabla_\mu X^\alpha = {1\ov N} \l( B^\mu \nabla_\mu X^\alpha
	- X^\mu \nabla_\mu B^\alpha \r) - K^\alpha_{\ \, \mu} X^\mu
	+ {X^\mu\ov N} \hat\nabla_\mu N \, n^\alpha \ . 
\ee
This formula can be used to express the derivative of $W^\alpha$ along
$n^\alpha$ which appear in the right-hand side of Eq.~(\ref{e:DV=...}). 
We can also apply it to $B^\alpha$ and get
\be \label{e:n.grad(B)}
 n^\mu \nabla_\mu B^\alpha = - 	K^\alpha_{\ \, \mu} B^\mu + 
	{B^\mu\ov N} \hat\nabla_\mu N \, n^\alpha \ .
\ee

By combining  Eqs.~(\ref{e:curl:def}), (\ref{e:DV=...}), 
(\ref{e:bar_eps=...})
and (\ref{e:n.grad(X)}), we arrive at the 3+1 decomposition of 
$(\nabla\wedge V)^\alpha$: 
\begin{eqnarray}
 (\nabla\wedge V)^\alpha & = & e^{-\Phi} \hat\epsilon^{\alpha\mu\nu} \l[
	N \hat\nabla_\mu W_\nu + {B_\mu\ov N} \l( B^\sigma \hat\nabla_\sigma
	W_\nu - W^\sigma \hat\nabla_\sigma B_\nu \r) 
	- Z {B_\mu\ov N} \hat\nabla_\nu N + B_\nu \hat\nabla_\mu Z 
	- 2 K_\mu^{\ \,\sigma} W_\sigma B_\nu \r] \nonumber \\
  & &	- e^{-\Phi} \ n^\alpha \  \hat\epsilon^{\lambda\mu\nu} B_\lambda
		\hat\nabla_\mu W_\nu  \label{e:rot(V),3+1} \ . 
\end{eqnarray}
The term with $\hat\epsilon^{\alpha\mu\nu}$ in factor is parallel 
to the hypersurface $\Sigma_t$ (because $\hat\epsilon^{\alpha\mu\nu}$ is),
whereas the second term is along $n^\alpha$.

Similarly Eqs.~(\ref{e:omega_def}), (\ref{e:v=exp_l}), (\ref{e:l=Nn-B}),
(\ref{e:bar_eps=...}) and (\ref{e:n.grad(B)}) leads to the
3+1 splitting of the rotation vector $\omega^\alpha$:
\be \label{e:omega,3+1}
  \omega^\alpha = e^{-2\Phi} \,
	\hat\epsilon^{\alpha\mu\nu} \l[ -{N\ov2} \hat\nabla_\mu B_\nu
	- B_\mu \hat\nabla_\nu N + K_\mu^{\ \,\sigma} B_\sigma B_\nu \r]
	+{e^{-2\Phi}\ov 2} \, n^\alpha \, 
	\hat\epsilon^{\lambda\mu\nu} B_\lambda
		\hat\nabla_\mu B_\nu .
\ee

We also need to perform the 3+1 decomposition of the second term on the
right-hand side of Eq.~(\ref{e:rotV,F}). The result is 
\be \label{e:Vxgrad(Phi),3+1} 
  \bar\epsilon^{\alpha\mu\nu} V_\mu \nabla_\nu \Phi = 
	e^{-\Phi}  \, \hat\epsilon^{\alpha\mu\nu}  \l[
	N W_\mu \hat\nabla_\nu \Phi + Z  B_\mu \hat\nabla_\nu\Phi
	- {B^\sigma\ov N} \hat\nabla_\sigma\Phi \, W_\mu B_\nu \r]
	- e^{-\Phi}  \, n_\alpha \, \hat\epsilon^{\lambda\mu\nu} B_\lambda
		W_\mu \hat\nabla_\nu \Phi  \ .
\ee

Thanks to Eqs.~(\ref{e:rot(V),3+1}), (\ref{e:omega,3+1}) and 
(\ref{e:Vxgrad(Phi),3+1}), 
the orthogonal projection of Eq.~(\ref{e:rotV,F}) onto the hypersurfaces
$\Sigma_t$ is straightforward and leads to the three-dimensional equation
\begin{eqnarray}
  & & \hat\epsilon^{ijk}  \l[ N \hat\nabla_j W_k 
	+ {B_j\ov N} \l( B^l \hat\nabla_l W_k - W^l \hat\nabla_l B_k \r)
	- Z {B_j\ov N} \hat\nabla_k N + B_k \hat\nabla_j Z
	-2 K_j^{\ \, l}\, W_l B_k \r] = 	 \nonumber \\
 & &\quad  \hat\epsilon^{ijk} \l[ e^{-\Phi}  \l( N \hat\nabla_j B_k +
	2 B_j \hat\nabla_k N - 2 K_j^{\ \, l} \, B_l B_k \r)  
  - N W_j \hat\nabla_k \Phi - Z B_j 
	\hat\nabla_k \Phi + {1\ov N} W_j B_k  B^l \hat\nabla_l \Phi  \r]
	+ e^\Phi \, F\, W^i  \label{e:rot(W)=...} \ .
\end{eqnarray}

The baryon number conservation equation~(\ref{e:baryon,cons}), once
recast in terms of three-dimensional quantities with the help of 
Eq.~(\ref{e:DV=...}), writes
\be \label{e:baryon,W}
  \hat\nabla_i W^i + W^i \hat\nabla_i \ln(N\Gamma n) +
	{Z\ov N} B^i \hat\nabla_i\ln(Z\Gamma n) = 0 \ .
\ee
A boundary condition on $W^i$ can be derived by multiplying this equation
by $n$ and setting $n=0$ (definition of the star's surface) into the 
result. One obtains, using Eq.~(\ref{e:Z(B,W)}),
\be \label{e:W_surf} 
	\l. \l( W^i - {B^i B_j W^j \ov N^2} \r) \hat\nabla_i \, n 
	 \r| _{\rm surface}
	= 0 \ .
\ee

The equations to be solved are the three-dimensional vector 
equation~(\ref{e:rot(W)=...}) and the scalar equation~(\ref{e:baryon,W}),
altogether with the boundary condition (\ref{e:W_surf}). 
Note that once Eq.~(\ref{e:rot(W)=...}) is satisfied, the other part of
the four-dimensional equation~(\ref{e:rotV,F}), 
namely the part along $n^\alpha$, is automatically fulfilled. Indeed,
the projection of Eq.~(\ref{e:rotV,F}) along $n^\alpha$ leads to
[cf. Eqs.~(\ref{e:rot(V),3+1}), (\ref{e:omega,3+1}),
(\ref{e:Vxgrad(Phi),3+1}) and (\ref{e:Z(B,W)})]
\be
   \hat\epsilon^{ijk} B_i \hat\nabla_j W_k = 
	e^{-\Phi} \hat\epsilon^{ijk} B_i \hat\nabla_j B_k
	- \hat\epsilon^{ijk} B_i W_j \hat\nabla_k \Phi
	+ {1\ov N} B_i W^i \, F \ ,
\ee
which is nothing else than the orthogonal 
projection of Eq.~(\ref{e:rot(W)=...}) onto $B^i$.

Referring to the discussion in Sect.~\ref{s:n_indep}, we conclude that
if a 3-vector $W^i$ of $\Sigma_t$ 
obeying Eqs.~(\ref{e:rot(W)=...}) and (\ref{e:baryon,W}) can be found,
the problem is completely resolved. Note that the baryon density $n$
which appears in Eq.~(\ref{e:baryon,W}) is given via the EOS by the 
log-enthalpy
$H$, itself being fully determined (for a fixed gravitational field)
by $W^i$ via the first integral of motion (\ref{e:intpre}), which becomes
in the counter-rotating case
\be \label{e:int_prem,contre-rot}
    H + \Phi + \ln\Gamma  = {\rm const.}\ , 
\ee
where [cf. Eqs.~(\ref{e:Lorentz}) and (\ref{e:V=W+Z})]
\be \label{e:Gam(W,Z)}
	\Gamma = (1- W_i W^i + Z^2)^{-1/2} \ .
\ee

\subsection{Formulation in terms of Poisson equations}

The equations to be solved, namely Eqs.~(\ref{e:rot(W)=...}) and
(\ref{e:baryon,W}), can be recast into Poisson equations by 
looking for solutions under the form
\be \label{e:W=A+psi}
	W^i = \hat\epsilon^{ijk} \hat\nabla_j A_k + \hat\nabla^i \psi \ ,
\ee
where $\psi$ is a scalar field and $A^i$ is a 
3-vector of $\Sigma_t$, which without any loss of generality can be taken to be
divergence-free :
\be
	\hat\nabla_i A^i = 0 \ .
\ee
This latter property implies that the curl of $W^i$ is related to 
the Laplacian of $A^i$ by
\be \label{e:rot(W)=Lap(A)}
   \hat\epsilon^{ijk} \hat\nabla_j W_k = - \hat\nabla_j \hat\nabla^j A^i
	+ \hat R^i_{\ \, j} A^j \ ,
\ee
where $\hat R_{ij}$ is the Ricci tensor of the 3-metric $h_{ij}$ of
$\Sigma_t$. 
Inserting Eq.~(\ref{e:rot(W)=Lap(A)}) into Eq.~(\ref{e:rot(W)=...})
leads to the following vector Poisson equation for $A^i$:
\begin{eqnarray}
  \hat\nabla_j \hat\nabla^j A^i & = & \hat\epsilon^{ijk} \Bigg\{
	- e^{-\Phi}  \l( \hat\nabla_j B_k +
	2 {B_j \ov N} \hat\nabla_k N - {2\ov N} K_j^{\ \, l} \, B_l B_k \r)  
  + W_j \hat\nabla_k \Phi + {1\ov N} \Bigg[ Z B_j 
	\hat\nabla_k \Phi - {1\ov N} W_j B_k  B^l \hat\nabla_l \Phi 
							\nonumber \\
 & & + {B_j\ov N} \l( B^l \hat\nabla_l W_k - W^l \hat\nabla_l B_k \r)
	- Z {B_j\ov N} \hat\nabla_k N + B_k \hat\nabla_j Z
	-2 K_j^{\ \, l} \, W_l B_k \Bigg] \Bigg\}
	- {e^\Phi \, F\ov N} \, W^i 
	+ \hat R^i_{\ \, j} A^j  \label{e:Lap(A)=...} \ .
\end{eqnarray}

The divergence of $W^i$ evaluated from Eq.~(\ref{e:W=A+psi}) is
\be \label{e:div(W)=Lap(p)+R}
	\hat\nabla_i W^i = \hat\nabla_i \hat\nabla^i \psi
		+ {1\ov 2} \hat\epsilon^{ijk} \hat R_{klij} A^l \ ,
\ee
where $\hat R_{klij}$ is the Riemann tensor associated with the 3-metric
$h_{ij}$. By virtue of the symmetry properties of the Riemann tensor in three
dimensions, the last term on the right-hand side of 
Eq.~(\ref{e:div(W)=Lap(p)+R}) vanishes identically, so that the divergence
of $W^i$ is simply the Laplacian of $\psi$ and Eq.~(\ref{e:baryon,W})
becomes
\be \label{e:Lap(psi)=...}
   \hat\nabla_i \hat\nabla^i \psi = - W^i \hat\nabla_i \ln(N\Gamma n) -
	{Z\ov N} B^i \hat\nabla_i\ln(Z\Gamma n) \ .
\ee

\subsection{Newtonian limit} \label{s:Newt_lim}
 
In the Newtonian limit, $\Phi$ takes the form (\ref{e:Phi_Newt}) and
$\omega^\alpha$ becomes [cf. Eq.~(\ref{e:omega=Omega})]
\be
 \omega^\alpha \stackrel{{\rm Newt.}}{\longrightarrow}  \vec\Omega_{\rm orb}	
	\ .
\ee

The rotating-coordinate shift vector $B^\alpha$ reduces to 
$B^\alpha = -\Omega_{\rm orb} \, m^\alpha$ [cf Eq.~(\ref{e:B:def})],
so that Eq.~(\ref{e:Lap(A)=...}) becomes 
\be
	\Delta \vec{A} = 2 \, \vec\Omega_{\rm orb} \ ,
\ee
whose divergence-free solution is
\be \label{e:A,newt}
   \vec{A} = {1\ov 2} \l( \vec{e}_z \times \vec{r} \r) ^2 \, 
	\vec\Omega_{\rm orb}	\ ,
\ee
where $\vec{e}_z := \vec\Omega_{\rm orb} / \Omega_{\rm orb}$. 
Finally, Eq.~(\ref{e:Lap(psi)=...}) becomes
\be
	\Delta \psi = - \vec{V}\cdot \vec{\nabla} \ln n \ .
\ee
Once this equation is solved, the fluid velocity field with respect
to the co-orbiting observer is computed by taking the
curl of (\ref{e:A,newt}) :
\be
  \vec{V} = - \vec\Omega_{\rm orb} \times \vec{r} + \vec{\nabla}\psi \ .
\ee
This is the solution obtained by Bonazzola \& Marck \cite{BoGHM92}. 

\section{Discussion} \label{s:discuss}

\subsection{Iterative method of resolution}

The resolution of the problem amounts to solving the vector Poisson equation
(\ref{e:Lap(A)=...}) for $A^i$ and the scalar Poisson equation 
(\ref{e:Lap(psi)=...}) for $\psi$, with the boundary condition 
(\ref{e:W_surf}) at the surface of each star. 
These equations involve Laplacian with respect to the {\em curved}
3-metric $h_{ij}$, so that even if the right-hand of the equations
is supposed to be known (e.g. from a previous step in an iterative method),
the numerical solution is not straightforward. A technique which has shown 
to be successful 
consists in introducing on $\Sigma_t$ a flat 3-metric $\bar h_{ij}$ and
decomposing the operators into flat-space
ordinary Laplacians plus curvature terms \cite{BoFGM96,BonFG97}. 

With this technique, the following iterative method can be
envisaged to get
counter-rotating binary neutron star configurations. The starting point
of the procedure can be very crude approximations such as constant density
spherical stars with $W^i=0$ and a flat spacetime metric.
At a given step, the gravitational field equations are to be 
solved\footnote{Although the gravitational field equations are not discussed
in the present paper, let us note that thanks to Eq.~(\ref{e:K=DB+DB}),
the momentum constraint equation can be expressed as a three-dimensional
vector Poisson equation for $B^i$.},
leading to new values for the functions $N$, $h_{ij}$, $K_{ij}$, $B^i$ and 
$\Phi$ [via. Eq.~(\ref{e:exp(2Phi)})].
The first integral of motion 
Eq.~(\ref{e:int_prem,contre-rot}) yields then to the value of the 
log-enthalpy $H$ throughout the stars. The Lorentz factor $\Gamma$
which appears in the first integral is to be evaluated by inserting the
previous step values of $W^i$ in Eq.~(\ref{e:Gam(W,Z)}). From $H$ the 
baryon number density $n$ is computed by means of the EOS
and inserted in the right-hand side of the scalar Poisson 
equation~(\ref{e:Lap(psi)=...}). In this right-hand side, 
as well as in the right-hand side of the vector Poisson equation 
(\ref{e:Lap(A)=...}), the value of $W^i$ is to be taken from the previous
step. This is of course also the case of the functions of $W^i$: $Z$
[deduced from $W^i$ by Eq.~(\ref{e:Z(B,W)})] and $F$ [deduced from $W^i$ by 
Eq.~(\ref{e:F:def})]. The Poisson equations (\ref{e:Lap(psi)=...}) and
(\ref{e:Lap(A)=...}) are then to be solved respectively for 
$\psi$ and $A^i$. The solutions are obtained up to the addition of harmonic
functions. These latter are determined in order that the boundary condition 
(\ref{e:W_surf}) is fulfilled. The vector field $W^i$ is deduced from the
obtained values of $\psi$ and $A^i$ via Eq.~(\ref{e:W=A+psi}) and a
new iteration may begin. 

Of course, we do not have any theorem about the convergence of this
iterative procedure. All that we can say is that similar schemes have been
applied successfully to the computation of axisymmetric \cite{BoGSM93}
and triaxial \cite{BonFG96,BonFG97} models
of single neutron stars and that in the axisymmetric case, a rigorous
proof of convergence has been recently given by Schaudt \& Pfister 
\cite{SchaP96}. 

\subsection{Conclusion}

Before the inner most stable orbit is reached, the evolution
of a binary system of neutron stars can be approximated by a sequence
of quasi-equilibrium configurations. For each of these configurations,
the spacetime possesses the helicoidal symmetry discussed in 
Sect.~\ref{s:quasi-equil}. 
The hydrodynamical part of the problem is then trivial
in the case of synchronized binaries, because of the existence of the 
first integral of motion (\ref{e:intpre_synchr}), which means that once
the gravitational field is known, the matter distribution in the
stars is obtained immediately. However realistic neutron star binaries
on the verge of coalescence are not synchronized but rather in counter-rotation. 
In this case, the velocity field inside the stars with respect to the
co-orbiting observer is not zero and has to be computed so that the
Euler equation (\ref{e:Euler,p}) is satisfied. 
We have presented a formalism which reduces the problem of finding this
velocity field to the resolution of three-dimensional scalar and vector
Poisson equations. We are currently applying the numerical techniques we 
have recently developed for solving such equations with spherical-type 
coordinates
\cite{BoFGM96,BonFG97} in order to get numerical models. We will present
the results of this work in a future article.

The formulation presented in this article is independent of any prescription
for solving the gravitational field equations (Einstein equations). It
simply relies on the assumption of the spacetime helicoidal symmetry
and can be used in conjunction with any set of equations for the
gravitational field, such as the Wilson \& Mathews' scheme 
\cite{WilsM89,WilMM96} or the 2-PN scheme recently proposed
by Asada \& Shibata \cite{AsadS96}. Note that both schemes involve nothing else
but the resolution of Poisson type equations, so that the method
that we propose
do not require a numerical technique specific to the hydrodynamical
equations.

\newpage

\end{document}